\documentclass[aps,prb, superscriptaddress, preprint, floatfix, longbibliography]{revtex4-1}

\usepackage{here}
\usepackage{graphicx}
\usepackage{wrapfig}
\usepackage{sidecap}

\usepackage{multirow}

\usepackage{amsmath}
\usepackage{amssymb}
\usepackage{braket}

\usepackage{hyperref}

\begin{document}

\title{The photoinduced transition in magnetoresistive manganites: a comprehensive view}
\date{\today}

\author{V. Esposito}
\affiliation{Swiss Light Source, Paul Scherrer Institut, 5232 Villigen PSI, Switzerland}

\author{L. Rettig}
\affiliation{Swiss Light Source, Paul Scherrer Institut, 5232 Villigen PSI, Switzerland}
\affiliation{Fritz Haber Institute, 14195 Berlin, Germany}

\author{E. Abreu}
\affiliation{Institute for Quantum Electronics, ETH Z\"{u}rich, 8093 Z\"{u}rich, Switzerland}

\author{E. Bothschafter}
\affiliation{Swiss Light Source, Paul Scherrer Institut, 5232 Villigen PSI, Switzerland}

\author{G. Ingold}
\affiliation{Swiss Light Source, Paul Scherrer Institut, 5232 Villigen PSI, Switzerland}
\affiliation{SwissFEL, Paul Scherrer Institut, 5232 Villigen PSI, Switzerland}

\author{M. Kawasaki}
\affiliation{RIKEN Center for Emergent Matter Science, Wako 351-0198, Japan}
\affiliation{Department of Applied Physics and Quantum Phase Electronics Center (QPEC), University of Tokyo, Tokyo 113-8656, Japan}

\author{M. Kubli}
\affiliation{Institute for Quantum Electronics, ETH Z\"{u}rich, 8093 Z\"{u}rich, Switzerland}

\author{G. Lantz}
\affiliation{Institute for Quantum Electronics, ETH Z\"{u}rich, 8093 Z\"{u}rich, Switzerland}

\author{M. Nakamura}
\affiliation{RIKEN Center for Emergent Matter Science, Wako 351-0198, Japan}
\affiliation{PRESTO, Japan Science and Technology Agency (JST), Kawaguchi, 332-0012, Japan}

\author{J. Rittman}
\affiliation{Swiss Light Source, Paul Scherrer Institut, 5232 Villigen PSI, Switzerland}
\affiliation{SwissFEL, Paul Scherrer Institut, 5232 Villigen PSI, Switzerland}

\author{M. Savoini}
\affiliation{Institute for Quantum Electronics, ETH Z\"{u}rich, 8093 Z\"{u}rich, Switzerland}

\author{Y. Tokura}
\affiliation{RIKEN Center for Emergent Matter Science, Wako 351-0198, Japan}
\affiliation{Department of Applied Physics and Quantum Phase Electronics Center (QPEC), University of Tokyo, Tokyo 113-8656, Japan}

\author{U. Staub}
\affiliation{Swiss Light Source, Paul Scherrer Institut, 5232 Villigen PSI, Switzerland}

\author{S. L. Johnson}
\affiliation{Institute for Quantum Electronics, ETH Z\"{u}rich, 8093 Z\"{u}rich, Switzerland}

\author{P. Beaud}
\email[Corresponding author: ]{paul.beaud@psi.ch}
\affiliation{Swiss Light Source, Paul Scherrer Institut, 5232 Villigen PSI, Switzerland}
\affiliation{SwissFEL, Paul Scherrer Institut, 5232 Villigen PSI, Switzerland}

\begin{abstract}
We use femtosecond x-ray diffraction to study the structural response of charge and orbitally ordered Pr$_{1-x}$Ca$_x$MnO$_3$ thin films across a phase transition induced by 800 nm laser pulses. By investigating the dynamics of both superlattice reflections and regular Bragg peaks, we disentangle the different structural contributions and analyze their relevant time-scales. The dynamics of the structural and charge order response are qualitatively different when excited above and below a critical fluence $f_c$. For excitations below $f_c$ the charge order and the superlattice is only partially suppressed and the ground state recovers within a few tens of nanosecond via diffusive cooling. When exciting above the critical fluence the superlattice vanishes within approximately half a picosecond followed by a change of the unit cell parameters on a 10 picoseconds time-scale. At this point all memory from the symmetry breaking is lost and the recovery time increases by many order of magnitudes due to the first order character of the structural phase transition.
\end{abstract}

\maketitle

\section{Introduction}
Non-thermal control of materials properties is key in developing technologically relevant applications. Notable examples are field-effect transistors and magnetic storage, where the properties of the device are controlled by an applied voltage or a magnetic field, respectively. Ultrashort light pulses have also been used to create non-thermal metastable states and this approach has been successfully applied to induce insulator-to-metal transitions \cite{Fiebig1998, Cavalleri2001, Tomeljak2009, Morrison2014} and to control superconductivity \cite{Fausti2011, Mankowsky2014, Mitrano2016}, ferroelectric polarization \cite{Rana2009, Mankowsky2017} or magnetic properties \cite{Finazzi2013,Kubacka2014, Auboeck2015, Baierl2016}. In these cases, however, the relaxation towards the ground state of the material does often not happen in a controlled manner and instead depends on intrinsic material properties. Understanding the underlying physics and time-scales of both the excitation process and the recovery mechanisms is thus of crucial importance to push toward potential technological applications.

The colossal magnetoresistance effect found in doped manganites is a well-known example of such a non-thermal effect. Magnetoresistive manganites exhibit a large variety of ordering phenomena as a function of doping and temperature due to the complex interplay between the charge, orbital, spin and structural degrees of freedom. At low temperature, two possible ground states are in competition: a ferromagnetic metallic phase and an antiferromagnetic, charge- and orbital-ordered insulating phase. The observed colossal magnetoresistance and the insulator-to-metal transition that occur in these materials are thought to be strongly related to this competition \cite{Dagotto2003}. Photo-induced transitions have been reported following the ultrafast excitation of the electronic system, with similar responses found when either the inter-atomic O \textit{2p} - Mn \textit{3d} e$_{g}$ charge-transfer resonance \cite{Polli2007} or the intra-atomic  Mn \textit{3d} transitions are excited \cite{Beaud2009}. This insulator-to-metal transition \citep{Miyano1997, Fiebig1998} is accompanied by a relaxation of the structural distortion \citep{Beaud2009, Caviezel2012} induced by the prompt suppression of the charge and orbital order \citep{Beaud2014}. The spin ordering is also significantly altered after photo-excitation. The antiferromagnetic order is melted \citep{Ehrke2011, Forst2011a, Zhou2014} and ferromagnetic correlations emerge \citep{Li2013, Lingos2017}. Other excitation mechanism such as resonant excitation of the lattice \cite{Rini2009, Forst2011a, Esposito2017}, pulsed magnetic fields \citep{Tokunaga1998}, intense x-ray radiation \cite{Kiryukhin1997,Garganourakis2012} or static illumination \cite{Elovaara2015} also dramatically alter the electronic, structural and magnetic properties of these manganites.

\begin{figure}
\centering
\includegraphics[scale=1]{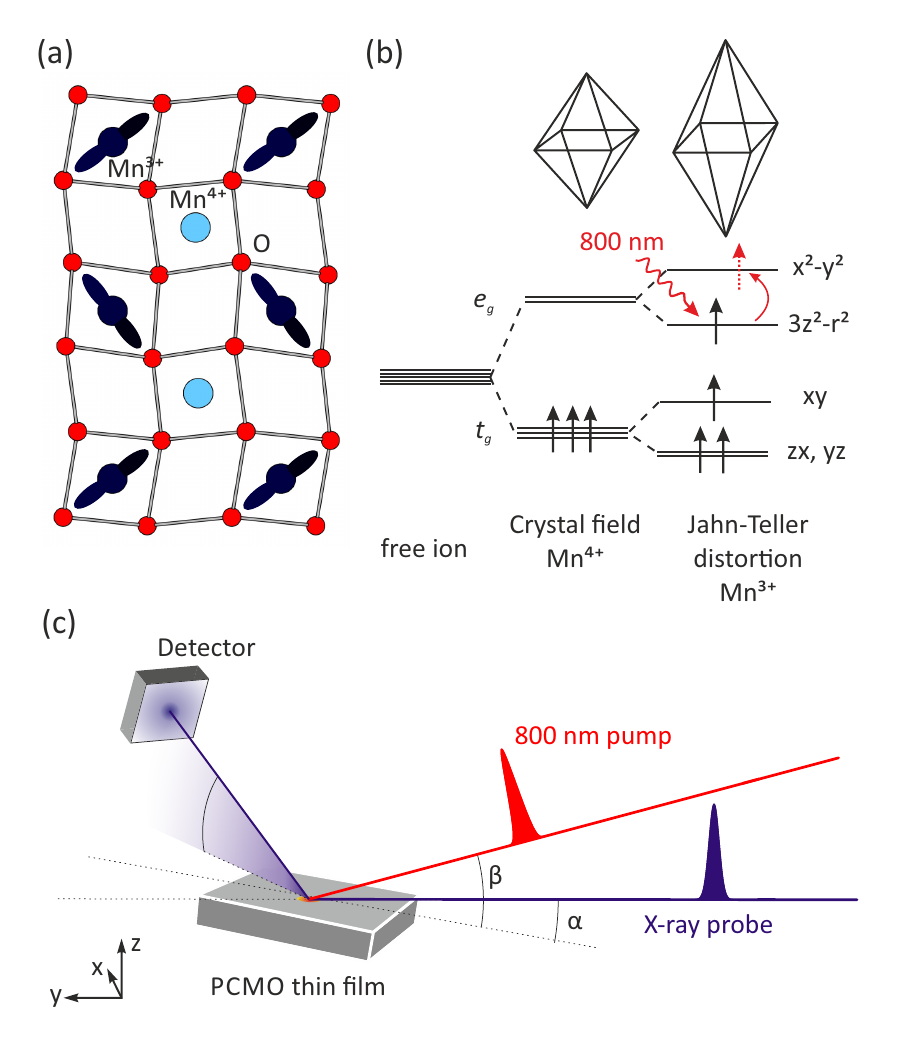}
\caption{(a) Charge and orbital order in PCMO in the \textit{ab}-plane. The apical oxygens and the rare-earth ions are omitted here for the sake of clarity. (b) Electronic structure of the \textit{d} orbitals of manganese as a function of the orbital occupancy and the corresponding distortion of the surrounding oxygen octahedra. The electronic transition induced by the 1.55 eV pump laser is indicated by the red arrow. (c) Sketch of the grazing incidence experimental setup for time-resolved x-ray diffraction. The incident angles $\alpha$ and $\beta$ are 0.5 and 10 degrees, respectively.}
\label{setup}
\end{figure}

In this Article we focus on the picosecond time-scale structural dynamics that follow the initial non-thermal melting of the superlattice in the three-dimensional charge and orbitally ordered manganite Pr$_{1-x}$Ca$_x$MnO$_3$ ($x=0.5,\ 0.4,\ 0.35$). The charge and orbital ordered (COO) state that emerges below T$_{\text{COO}}$ consists of a checker-board pattern of Mn$^{3+}$ and Mn$^{4+}$ in the \textit{ab}-plane and the alignment of the  Mn$^{3+}$ $3z^2-r^2$ orbitals in a zig-zag pattern along the \textit{Pbnm} $b$-axis. This first order transition induces a Jahn-Teller (JT) distortion on the Mn$^{3+}$ sites, leading to a lowering of the crystal symmetry from orthorhombic $Pbnm$ to monoclinic $P2_1/m$ \cite{Goff2004}. The COO state and a sketch of the Mn electronic configurations are shown in Fig. \ref{setup} (a) and (b). Because of the doubling of the unit cell along the crystallographic $b$-axis, this transition is directly observable in x-ray diffraction experiments by the appearance of superlattice (SL) reflections of the type $(h \frac{k}{2} l)$ \cite{Zimmermann1999}$^,$ \footnote{All the Miller indices in this paper are given with respect to the high temperature orthorhombic cell.}. A second order transition to a CE-type antiferromagnetic order occurs at the Neel temperature T$_{\text{N}} < \text{T}_{\text{COO}}$ \cite{Dagotto2003,Staub2009}.

Our study discusses not only the entire phase transition to the higher symmetry state but also the recovery back to the initial lower symmetry state. Together with the result of previous studies this investigation proposes a unified picture for this photo-induced non-equilibrium phase transition.

\section{Experimental details}
The 40 nm thick PCMO thin films were grown by pulsed laser deposition on $(110)$-oriented  (LaAlO$_3$)$_{0.3}$–(SrAl$_{0.5}$Ta$_{0.5}$O$_3$)$_{0.7}$ substrates \cite{Okuyama2009}. Three samples with different doping concentration ($x=0.5, \ 0.4, \text{ and } 0.35$) were investigated in this work. The COO transition temperatures were characterized by transport measurement yielding T$_{\text{CO}} = 221, 206 \text{ and } 190$ K respectively. In the diffraction experiments reported here the sample was cooled down to 100 K with a nitrogen cryoblower.

The time-resolved x-ray experiments were carried out at the FEMTO slicing source at the Swiss Light Source \cite{Beaud2007}. This source produces pulses of about 140 femtoseconds duration from electrons that are modulated by a short laser pulse in the storage ring. A sketch of the experimental setup is shown in Fig. \ref{setup} (c). Using a double Mo/B$_{4}$C multilayer monochromator the x-ray energy was set to 7 keV, a good compromise between the optimal source flux and minimum air-transmission losses.
The penetration depth of the pump and the probe were matched by a small 0.5 degree incident angle of the x-ray beam and the footprint on the sample was minimized by small vertical focus of the beam of 10 $\mu$m \cite{johnson2008}. The beam remained unfocussed horizontally with a width of about 300 $\mu$m. The sample was excited with weakly-focused ($520 \times 620$ $\mu$m$^2$) \textit{p}-polarized 100-fs pulses at 800-nm wavelength entering the sample at 10 degrees grazing incidence. The angle mismatch between the pump and the probe increases the time resolution to approximately 150 fs. The repetition rates of the optical pump and the x-ray probe were 1 and 2 kHz respectively. In this way the probed signal alternates between pumped and unpumped states of the sample, allowing to compensate for slow drifts in the x-ray source, pump laser, detector, and electronics \cite{Saes2004}.

The diffracted intensity of superlattice peaks was measured with an avalanche photodiode (APD), providing enhanced sensitivity for weak signals. Instead the Pilatus pixel detector \citep{Henrich2009} was used in gated mode to investigate the more intense regular Bragg reflections. Both detectors provide a photon-counting operation mode that is well-suited for these low flux diffraction experiments. In addition the pixel detector allows to map the peak displacements in reciprocal space, disentangling in-plane and out-of-plane motions. The x-ray energy is above the Mn $K$ edge and therefore x-ray fluorescence contributes significantly to an homogeneous background. For the data taken with the 2-dimensional detector, the background is given by a featureless region of interest in the images, while in the case of the APD point detector, it is estimated from the constant baseline of the rocking curve scans.

\section{Results and analysis}
Our results are discussed in three subsections, which correspond to the dynamics at different timescales. First the subpicosecond structural dynamics is examined. On this time-scale, the atomic motions are limited to isochoric dynamics, involving small atomic motion carried by coherent optical phonons that do not alter the dimensions of the underlying unit cell. In a second section, we investigate the picosecond dynamics of the lattice and reveal a second step of the structural transition. This slower component of the phase transition includes a change of the crystallographic unit cell from the low temperature monoclinic structure to the high temperature orthorhombic cell. Finally, we discuss the recovery to the COO ground state.

\subsection{The ultrafast isochoric transition}
\label{dopdep}
\begin{figure*}
\centering
\includegraphics{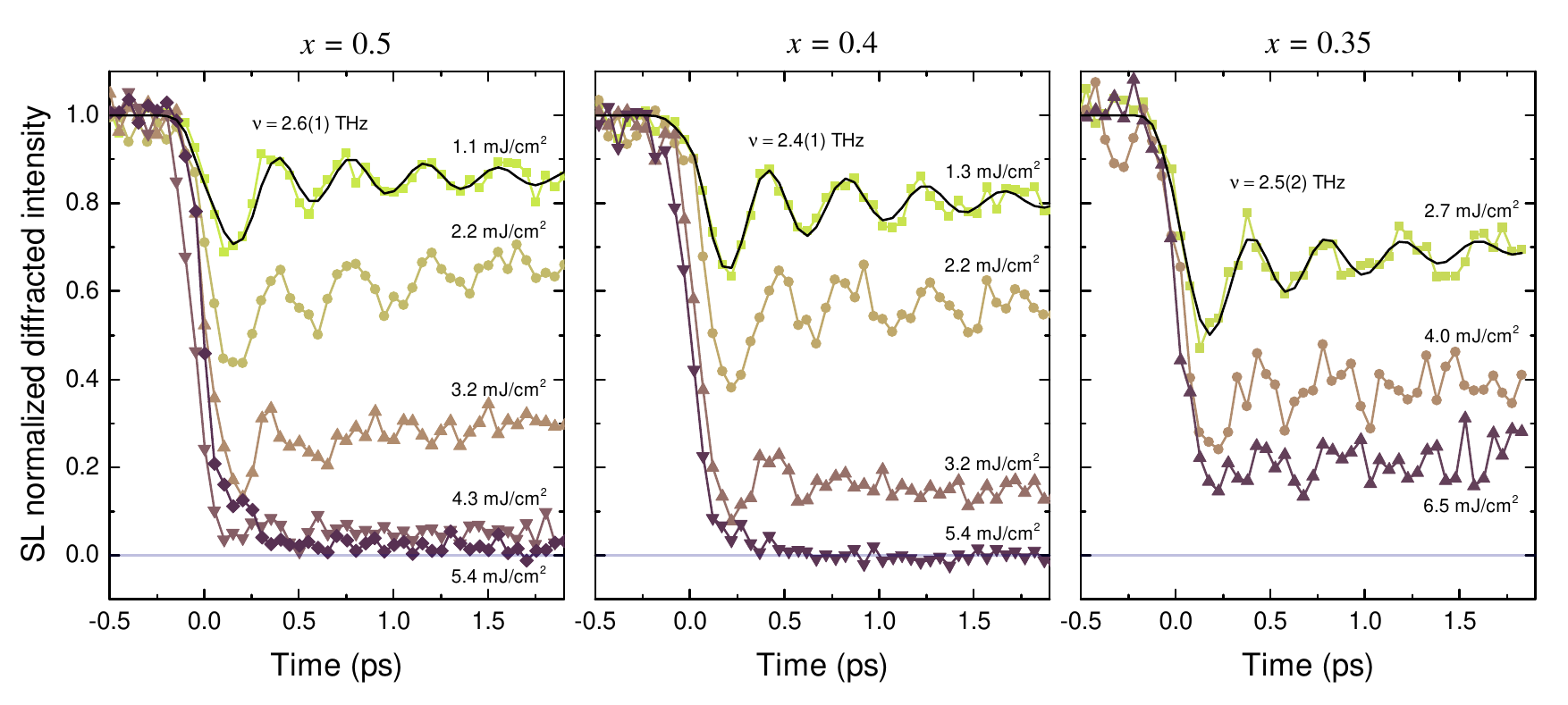}
\caption{Time-dependence of the diffracted intensity of the $(\bar{2} \bar{\frac{3}{2}} 0)$ superlattice reflection at 100K for different pump fluences and doping concentrations $x$. The black lines are fits used to extract the frequency $\nu$ of the observed coherent oscillation.
}
\label{SL1_tdep}
\end{figure*}

An overview of the fluence dependence of the dynamics of the $(\bar{2} \bar{\frac{3}{2}} 0)$ superlattice reflection of the PCMO films  is given in Figure \ref{SL1_tdep} for three doping concentration $x=0.5, 0.4, \text{ and } 0.35$. This type of reflections emerge as a direct consequence of the doubling of the unit cell below the ordering temperature. Their structure factors thus provide direct information on the magnitude of the low temperature structural distortion. All three doping concentrations display qualitatively similar time-dependences following photo-excitation: a fast drop in intensity for all fluences and a clear prominent coherent $2.5$ THz oscillation at low and intermediate fluences. The data for the $x=0.5$ and $x=0.4$ samples are very similar and the peak intensity disappears in both cases for fluences above about $5$ mJ/cm$^2$.

\begin{figure}
\centering
\includegraphics[scale=1]{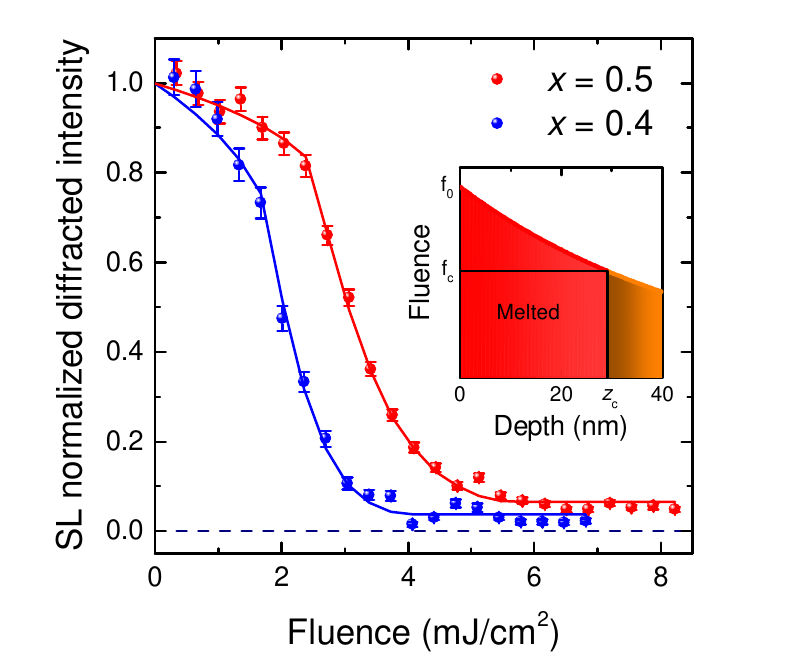}
\caption{Fluence scans of the SL peak 20 ps after excitation for the doping concentration $x=0.5$ (red) and $x=0.4$ (blue). The lines are fits with a Landau-like order parameter function (cf main text). The inset shows an excitation profile for a surface fluence $f_0$ between $f_c$ and $f_c'$, for which the bottom layers of the film experience a fluence below $f_c$.}
\label{wp_doping}
\end{figure}

For the $x=0.5$ sample, the dynamics and critical behaviour are consistent with a previous study of the charge order and lattice dynamics \citep{Beaud2014}. The superlattice peak disappears at times $t > 1$ ps for the same effective critical fluence $f_c' \approx 5$ mJ/cm$^2$. This corresponds to the incident fluence at which the deepest layers of the film are exposed to the critical fluence necessary to melt the charge order, i.e. when the whole film has undergone the phase transition \cite{Beaud2014}. Taking into account the inhomogeneous excitation profile, this corresponds to a true critical fluence $f_c \approx 2.4$ mJ/cm$^2$, the absorbed fluence at which the COO melts at the surface of the film. Between $f_c$ and $f_c'$ there is a coexistence of melted and non-melted layers (inset Fig. \ref{wp_doping}).

The critical fluences can be also fitted in the fluence-dependent drop of the superlattice intensity at 20 ps (Fig. \ref{wp_doping}). In equilibrium the structure factor of the superlattice reflection is an order parameter of the structurally distorted phase. Out of equilibrium this is however not always the case, as the transient structure does not necessarily corresponds to a minimum of the potential energy surface. In our experiment, the oscillatory dynamics are gone after a few picoseconds, suggesting that the dynamical component of the response has damped out and the structure is effectively tracking the position of a minimum in the potential energy surface. The structure factor $F$ of the reflection is thus modeled by a Landau-like order parameter, constructed by splitting the film into N layers \cite{Beaud2014}:

\begin{equation}
F = \frac{1}{N} \sum_i^N \left( 1- \frac{f(z_i)}{f_c} \right)^{\gamma} \text{,}
\label{Landau_model}
\end{equation}
where 
\begin{equation}
f(z_i) = f_0 e^{-z_i/z_0}
\label{BeerLaw}
\end{equation}
is the fluence seen by a layer at depth $z_i$ in the sample, $z_0=48$ nm is the effective penetration depth, accounting for the geometry of the experiment. The critical fluence $f_c$ and the critical exponent $\gamma$ are fitted to the data in Fig. \ref{SL1_fludrop}. A small scaling factor is used to account for the fact that the reflection does not vanish at the highest fluences. The extracted critical fluences are $f_c=2.4(2)$ mJ/cm$^2$ for the $x=0.5$ sample and $f_c=1.7(1)$ mJ/cm$^2$ for the $x=0.4$ sample. The fitted critical exponents are $\gamma = 0.17(4)$ and $\gamma = 0.27(6)$, consistent with similar measurements \cite{Beaud2014}.

In the $x=0.35$ sample the reflection does not disappear even for the highest fluences. This may, however, be the result of large metal contacts present on the sample surface, which might have partially shadowed the excitation pulse in the grazing incidence geometry used for the experiment. This shadowing affects both the x-ray probe pulse and the optical pump. The overall Bragg peak intensity decreases and the inhomogeneously distributed pump intensity reduces the fluence dependent drop and the phonon amplitude, making the extraction of quantitative parameters such as the critical fluence unreliable. Putting aside these measurements artifacts, the overall dynamic response is qualitatively very similar for the different doping concentrations and we will therefore focus on the following on the half-doped $x=0.5$ film only.

\subsection{Monoclinic to orthorhombic transition}
\label{mono_ortho}
In the adiabatic limit, heating the system above T$_{\text{COO}}$ includes a change from a monoclinic to orthorhombic lattice in addition to the removal of the superlattice. However, such changes cannot occur on the fast time scales that were discussed in the previous section. We thus expect that on longer time scales the unit cells should deform to an orthorhombic shape. Superlattice reflections are not adequate to study the changes of the unit cell across a phase transition, since they vanish once the change of superlattice symmetry has occurred. Therefore we chose to measure the $(002)$ Bragg reflection, whose structure factor is only weakly altered by the formation of the superstructure, and can be monitored throughout the transition.

\begin{figure*}
\centering
\includegraphics[scale=1]{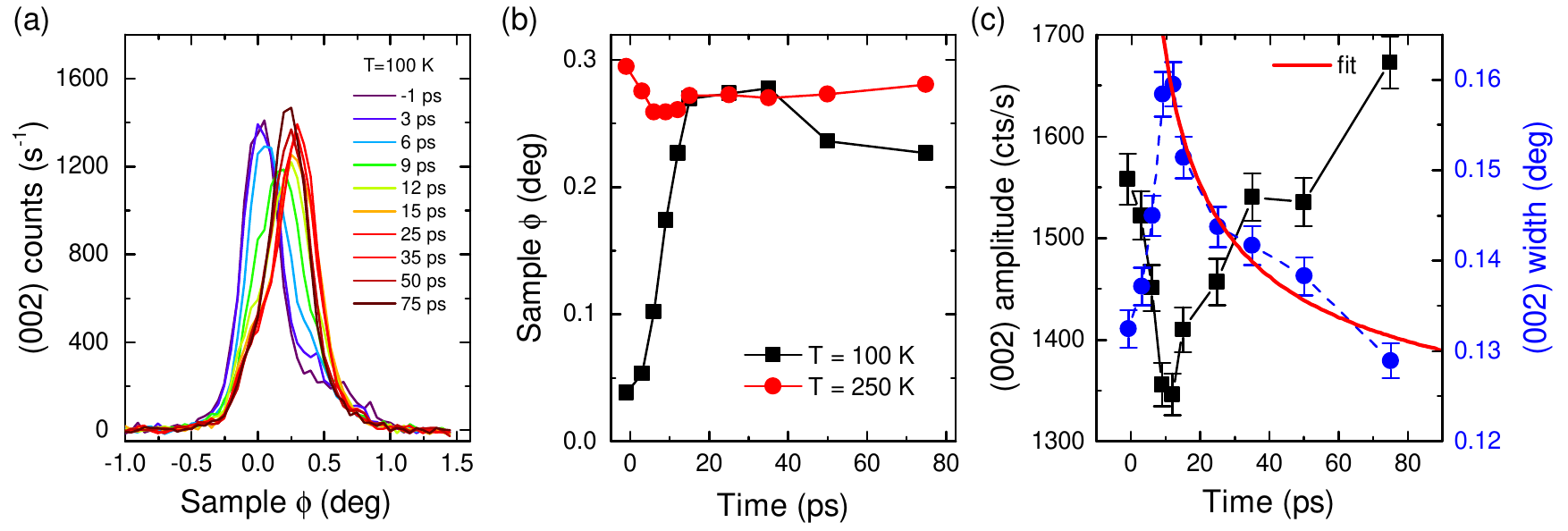}
\caption{
(a) Time dependence of the $(002)$ Bragg reflection at a fluence of 8.2 mJ/cm$^2$ ($T=100$ K). The $\phi$ axis indicates the sample rotation around its surface normal and the reference ($\phi$ = 0) is set to the low temperature peak position.
(b) Fitted $(002)$ position as a function of time at 100 K (black squares) and 250 K (red dots). The error bars (not shown) are about the size of the symbols.
(c) Amplitude and width of the $(002)$ peak at $T=100$ K. The fit to the peak width for times later then 10 ps (red line) is discussed in the main text.
}
\label{002_tdep}
\end{figure*}

The time dependence of the $(002)$ reflection at 100 K, i.e. below T$_\text{COO}$, is summarized in Fig. \ref{002_tdep}. In panel (a), the intensity of the reflection is plotted as a function of the sample rotation $\phi$ for various time delays at an excitation fluence $f=8.2$ mJ/cm$^2$, fully sufficient to melt COO and the superlattice in the entire probed volume. An equivalent dataset (not shown) was taken above T$_{\text{COO}}$ to compare the induced dynamics in the absence of the structural phase transition. The shoulder around $\phi=0.5$ degree rotation in Fig. \ref{002_tdep} (a) matches the high temperature peak position and is attributed to a few strained layers and/or defects, probably near the film/substrate interface, that do not undergo the transition when cooling the sample. Its large width indicates indeed a small correlation length, as expected from diffraction by a very thin layer.

The peaks were fitted with a Lorentzian and the extracted parameters are summarized in panels (b) and (c) of Figure \ref{002_tdep}. The Bragg peak position versus time is shown in Figure \ref{002_tdep} (b) for temperature above and below T$_{\text{COO}}$. In the high temperature phase, the peak position is almost constant as a function of time with small variations due to lattice heating dynamics and a strain wave traversing the thin sample. In the low temperature phase, however, the peak displacement is substantial. The Bragg peak moves to its high temperature position within approximately 20 ps demonstrating that the unit cell is driven within this short time scale away from its monoclinic structure and towards the high temperature orthorhombic one.

The time dependence of the peak width and amplitude are shown in Fig. \ref{002_tdep} (c). The clear correlation between these two parameters results in an approximately constant integrated intensity throughout the transition. Hence the structure factor is not significantly altered in the process and its variation can be neglected. The increase in peak width until about 15 ps indicates a decrease in correlation length, probably due to the coexistence of orthorhombic and monoclinic domains. The initial correlation length and amplitude recover within the subsequent 60 ps.

\begin{figure}
\centering
\includegraphics[width=8.4cm]{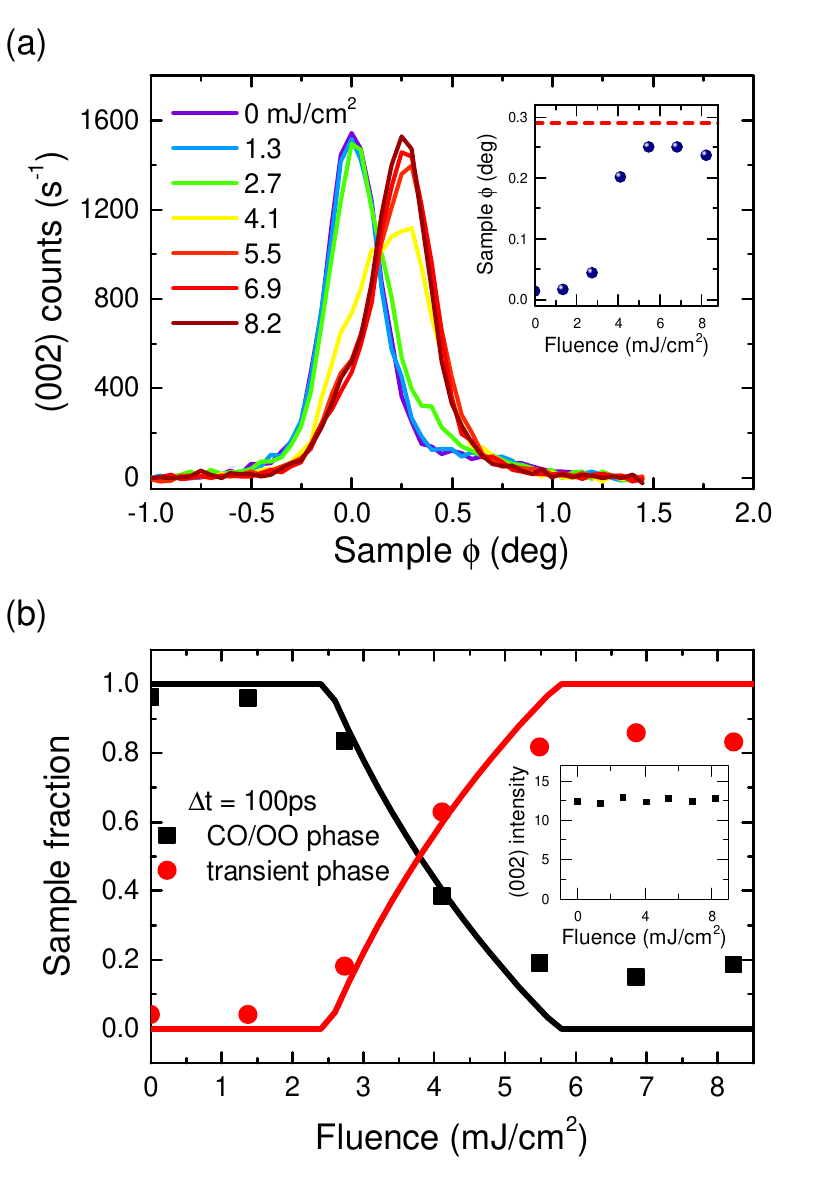}
\caption{
(a) Fluence dependence of the $(002)$ regular Bragg reflection 100 picoseconds after excitation. The inset shows the fitted peak position as a function of fluence. The red dashed line indicates the position of the high-temperature peak taken from Fig. \ref{002_tdep} at $t=-1$ ps.
(b) Fractions of each phases as a function of fluence 100 ps after excitation. They are extracted by fitting the data with a sum of two peaks whose centers are fixed at the peak position of the low and high temperature phases (symbols). The model (lines) takes into account the inhomogeneous excitation profile in the sample (see text for details). The inset shows the integrated intensity of the peak as a function of fluence.
}
\label{002_fludep}
\end{figure}

The fluence dependence for the $(002)$ reflection at a fixed time-delay of 100 ps is shown in Fig. \ref{002_fludep}. The peak is also fitted with a Lorentzian and the extracted peak position as a function of fluence is shown in the inset. Clear changes can be seen at $f \approx 2.4$ mJ/cm$^2$ and $f \approx 5$ mJ/cm$^2$, corresponding closely to the two critical fluences $f_c$ and $f_c'$ introduced in the previous section. For fluences below $f_c$ the peak position remains fixed. Above $f_c$ the peak gradually shifts and stabilizes approximately at the high temperature position for $f > f_c'$. Similar to the time-dependence, the fluence dependence of the peak width and amplitude are anti-correlated, keeping the integrated intensity approximately constant (see inset Fig. \ref{002_fludep} (b)).

As stated in the previous section, the fluence $f_c$ corresponds to the critical fluence necessary to melt the COO at the surface \citep{Beaud2014}, indicating that the monoclinic-to-orthorhombic transition only occurs once the COO has entirely disappeared. This sharp threshold behaviour is somewhat smeared between $f_c$ and $f_c'$ because of the inhomogeneous excitation profile.

In Figure \ref{002_fludep} (b), we compare the sample fraction in each phase with a simple binary model where we account for the excitation gradient \citep{Moehr-Vorobeva2011}. To extract the volume fraction, two peaks are fitted to the data. The peak positions are constrained to the low and high fluence position respectively, each peak being a measure of one of the phases, and the widths are assumed to be constant. The volume fraction of each phase is then given by the integrated intensity of its corresponding peak, normalized to the total intensity. The wide shoulder around $\phi=0.5$ in the unpumped trace is taken as an additional background constant for all fluences.

We also calculate the critical depth $z_c$ by inverting Eq. \ref{BeerLaw}
\begin{equation}
z_c = z_0 \ln(f_c/f_0) \text{,}
\end{equation}
and calculate for each fluence the volume fraction below and above $f_c$. This simple model describes the data well (lines in Fig. \ref{002_fludep} (b)), demonstrating the threshold nature of the transition. The discrepancies at high fluence are attributed to the small changes of the structure factor and the partial recovery that starts already at about 50 ps after excitation (Fig. \ref{002_tdep} (b)).

\subsection{Recovery dynamics}
\label{reco_dyn}
\begin{figure}
\centering
\includegraphics[scale=1]{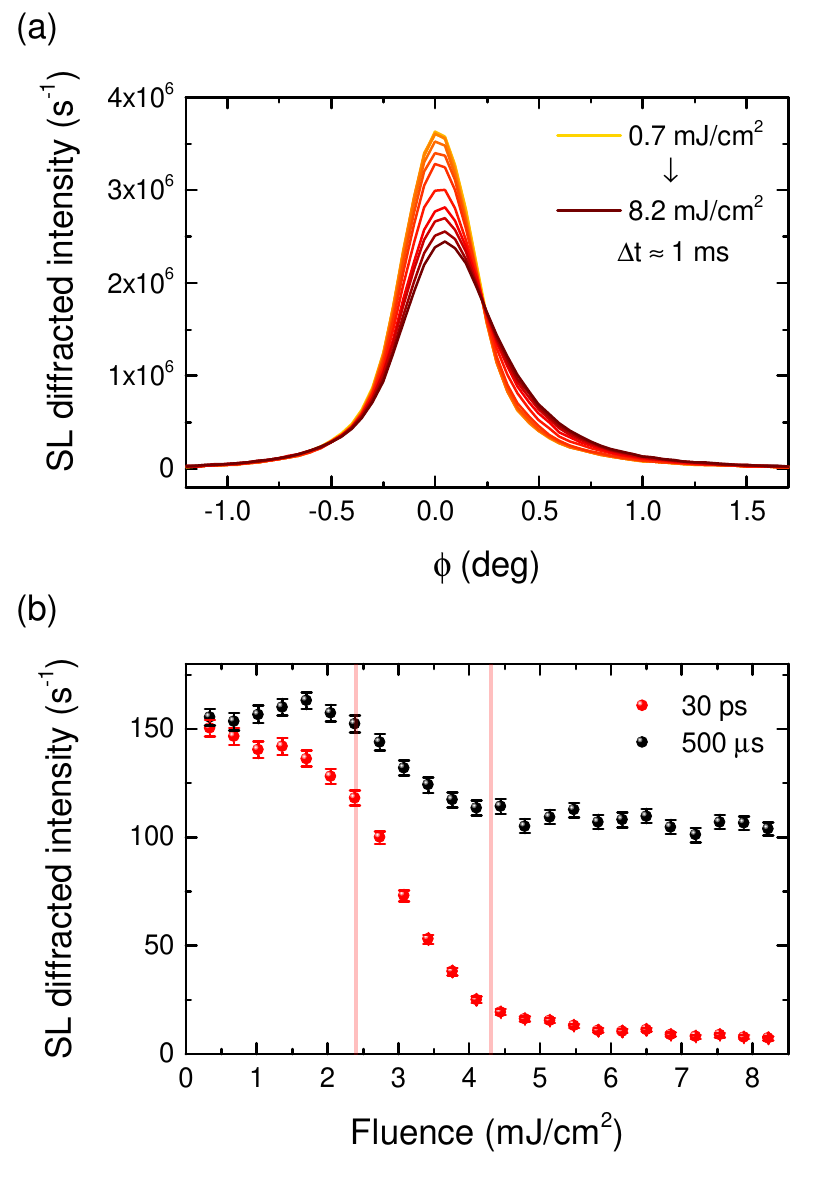}
\caption{
(a) Rocking curves scans of the $(\bar{2}\bar{\frac{3}{2}}0)$ superlattice peak measured for various pump fluences just before the arrival of the pump pulse. These data were taken at 1 kHz using 70 ps duration x-ray pulses (slicing mode off), explaining the much higher intensity.
(b) Fluence dependence of the $(\bar{2}\bar{\frac{3}{2}}0)$ reflection at $\phi=0$ and 30 ps (red), respectively 500 us (black) after excitation. The 500 $\mu$s trace is usually taken as the unpumped reference for the pump-probe experiment at the FEMTO slicing source.
}
\label{SL1_fludrop}
\end{figure}

In a pump-probe experiment the measured time dependent variable is usually normalized to the unperturbed value of the system before arrival of the pump, working under the assumption that the system fully recovers between subsequent pump pulses. Hence most studies concentrate on the pump induced relative changes (see Fig. \ref{SL1_tdep} for example). In Fig. \ref{SL1_fludrop} (a), we plot rocking curves of the $(\bar{2}\bar{\frac{3}{2}}0)$ SL reflection for a set of fluences taken just before the arrival time of the next pump, corresponding to a delay of almost $1$ ms after the previous pump pulse. It is clear that the system at higher pump fluences does not fully return to the ground state. This reduction is not due to permanent damage because the intensity fully recovers when blocking the pump. The broader width at the highest fluences indicates smaller domain sizes and the slight shift possibly a marginal increase of the base temperature. The integrated intensity overall decreases by about $15 \%$ at the highest fluences. Considering that the integrated intensity of the $(002)$ reflection remains constant as a function of fluence (inset of Fig. \ref{002_fludep} (b)), we conclude that these changes originate exclusively from smaller ordered domains, which results in an increased number of domain walls that do not contribute to the superlattice reflection.

Figure \ref{SL1_fludrop} (b) shows the maximum intensity of the $(\bar{2}\bar{\frac{3}{2}}0)$ reflection as a function of fluence $30$ ps and 500 $\mu$s after excitation. The intensity at $t=30$ ps shows the expected kink at $f_c$ and vanishes around $f_c'$ when the COO is melted throughout the film. The unpumped reference ($500$ $\mu$s curve, black circles) remains stable until the critical fluence is reached and then continuously drops until $f_c'$. For larger fluence it again remains constant at about $70\%$ of its equilibrium value. This decrease in intensity is only partially compensated by the increased width (Fig. \ref{SL1_fludrop} (a)).

Upon the observation of such long-lived dynamics, the pump-probe repetition rate ideally would have to be reduced \citep{Zhou2014}.  This was unfortunately not feasible here due to the very low flux of the slicing source.
Although we should be careful treating data at fluences above $f_c$, there is no indication that the phase after 1 ms is different from the original COO phase. We only observe a small reduction of the domain size and a slight increase of the average temperature. With this in mind, we argue that we can still discuss some aspects of the critical behavior and the recovery dynamics.

\section{Discussion}
The subpicosecond dynamics resulting from melting the superlattice in the ground state of mixed valence manganites with an ultrashort 800 nm laser pulse have been extensively discussed in previous works \citep{Beaud2009, Caviezel2012, Caviezel2013, Beaud2014}. Our current results are consistent with these studies and here we simply summarize the main conclusions. The ultrashort $1.5$ eV excitation  effectively removes an electron from the elongated $3z^2-r^2$ orbital of the Mn$^{3+}$ $e_g$ band and transfers it into the more symmetric $x^2-y^2$ orbital (Fig. \ref{setup} (b)). This initial step instantaneously decreases the orbital order and is followed by a rapid delocalization of the excited $e_g$ electrons, thereby suppressing the charge order \citep{Beaud2014}. These rapid changes of the electronic properties launch an ultrafast structural phase transition. The fast suppression of COO releases the Jahn-Teller distortions and the superlattice collapses. The sudden collapse drives coherent vibrational motion along several optical $A_g$ phonon mode coordinates \citep{Matsuzaki2009}, ranging in frequency from the Jahn-Teller mode at ~14.5 THz to a slow vibrational mode at ~2.5 THz that quickly dominates the dynamical response \citep{Lim2005, Caviezel2012}.  This last frequency is believed to correspond to a 2.7 THz mode observed in equilibrium and attributed to the motion of the rare earth Pr/Ca ions\cite{Amelitchev2001}. It was shown, however, that the atomic motion corresponding to the observed 2.5 THz oscillation cannot be exclusively attributed to motion of these ions \citep{Caviezel2013}. The observed oscillation is rather the result of the renormalization of the excited modes, due to nonlinear phonon coupling \cite{Beaud2014}. For low excitation, the electronic order and structural distortions are only partially  relaxed. But at sufficiently high fluences, the COO is completely destroyed and the potential energy surface shifts towards a high symmetry point, as shown by the complete disappearance of superlattice reflections within $0.5$ ps (Fig. \ref{SL1_tdep}) \citep{Caviezel2012, Beaud2014}.

The additional holes added when moving away from the optimum half-doped system are often described as defects within the COO pattern \cite{Dagotto2003}. This explains well the lower critical temperature observed for smaller x values, and similarly a lower critical fluence is expected for the photo-induced transition. A lower critical fluence is indeed found for the $x=0.4$ sample but no quantitative estimation of the critical fluence can be made for the $x=0.35$ sample due to the unknown excitation profile. We would, however, expect it to be even lower. Overall it is clear that the dynamics are very little affected by the doping concentration and mainly differ in their respective critical fluences.

\begin{figure}
\centering
\includegraphics[scale=1]{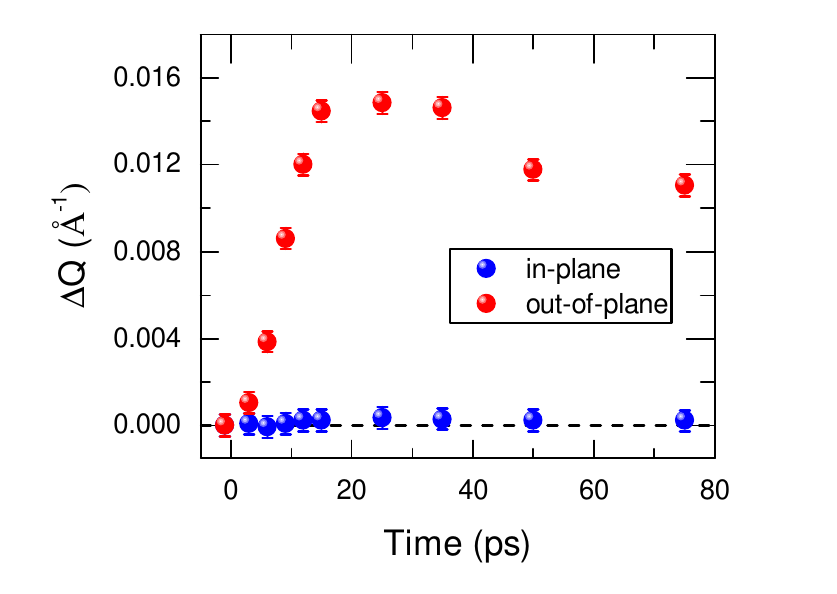}
\caption{Decomposition of the (002) peak position along the out-of-plane $(\bar{1} \bar{1} 2)$ and in-plane projections. These data are extracted from the reciprocal space analysis of the data in Fig. \ref{002_tdep}, taking advantage of the 2-dimensional Pilatus pixel detector.}
\label{002_compproj}
\end{figure}

The second step of the structural phase transition, the monoclinic-to-orthorhombic transition, takes place only for fluences exceeding a critical fluence $f_c$, when the COO and the superlattice are completely destroyed. As shown by the dynamics of the $(002)$ reflection (Fig. \ref{002_tdep}), this step is much slower than the disappearance of the superstructure because it involves changes of the lattice constants which are carried by acoustic phonons and consequently limited by the speed of sound. As a matter of fact on these fast time scales only changes of the out-of-plane lattice constants occur whereas the in-plane lattice constants remain locked to the substrate \citep{Okuyama2009}. Indeed the decomposition of the Bragg peak position along the $(\bar{1} \bar{1} 2)$ out-of-plane and in-plane directions in Fig. \ref{002_compproj} shows  only a motion along the out-of-plane direction. The time-scale on the order of 10 ps (Fig. \ref{002_tdep} (b)), is consistent with the time required for an acoustic strain wave to propagate at the speed of sound $v_s \approx 5$ nm/ps \citep{Gaidukov1998} across the 40 nm thickness of the film.

After the peak has reached the high temperature position, its intensity and width still evolve and relax back to their equilibrium value in the next $50$ ps (see Fig. \ref{002_tdep} (c)). This relaxation is well described by a $t^{-1/2}$ dependence of the peak width (red line in Fig. \ref{002_tdep} (c)), stemming from a $\sqrt{t}$ dependence of the domain size, which is predicted for domain growth dynamics following the quench of the order parameters \citep{Grest1988,Chen1994,Laulhe2017}. It indicates that the high symmetry phase grows following a nucleation process. In particular the integrated intensity is constant showing that the growth of the correlation length is not due to an increase of the transient phase fraction but is rather due to domain wall dynamics, which promotes larger domain size. The observation of both the $\sqrt{t}$ dependence of the domain size and a threshold as a function of fluence points toward a first-order-like description of this second step of the structural transition, while the initial COO melting and the subsequent collapse of the superlattice are continuous and were modeled with a single Landau-like order parameter \citep{Beaud2014}.

These dynamics on a time scale of tens of picoseconds round up our understanding of the laser-induced phase transition: we have now identified three steps, each characterized by their relevant time-scales. First, the charge and orbital order is destroyed by the photo-excitation of the electrons ($\tau_1 < 80$ fs) \citep{Beaud2014}. Second, the Jahn-Teller distortion collapses triggering the rearrangement of the unit cell in the high symmetry configuration ($\tau_2 \approx 0.5$ ps) \citep{Beaud2009, Caviezel2012, Beaud2014}. Finally, within $10-20$ ps this non-equilibrium state relaxes towards a thermodynamic metastable state, corresponding to the high temperature orthorhombic unit cell. In the adiabatic case these different components are indissociable, but they become separated in the time domain, as the change of each component is mediated by dynamics with different time scales. After this last step, the material is structurally in a configuration very close to that of the high temperature phase. The electronic and transport properties are, however, quite different. A metallic metastable state \cite{Fiebig1998} and the rapid formation of ferromagnetic ordering \cite{Li2013} has been observed, differing significantly from the thermodynamic state above T$_{COO}$.
 
The ferromagnetic correlations are found to emerge within the first 120 fs after excitation \citep{Li2013, Lingos2017} and the antiferromagnetic magnetic order is suppressed on a time-scale faster than 250 fs \cite{Forst2011a}. These time-scales are limited by the experimental resolutions and are compatible with the fast reduction of the COO that we observe. Interestingly, a fast rise of the magnetization is only found above a threshold fluence $f_{th} = 2.5$ mJ/cm$^2$, which matches surprisingly well the critical fluence $f_c$ found here for the COO melting and the structural phase transition. Below this threshold, the ferromagnetic volume increases on a slower picosecond time-scale, assigned to thermal effects. Outside the COO region of the phase diagram, manganites often develop a ferromagnetic or canted ordering at low temperature, which, in association with the double exchange mechanism, leads to the observed metallic behaviour \cite{Dagotto2003}. Due to the small ionic size of Pr and Ca, the thermodynamic ferromagnetic phase in PCMO is, somewhat unusually, insulating. In cuprates the charge ordered state was shown to hamper the superconducting dome \cite{Chang2012, SilvaNeto2013}. Similarly here, the ferromagnetic metallic state may thus be hindered by the COO and may become a hidden metastable state that is only accessible by applying a magnetic field for example or by non-thermal melting of the COO via ultrafast photo-excitation.

The completion of the transition above $f_c$ significantly modifies the relaxation dynamics. For example, the time required to recover the COO superlattice increases by many orders of magnitude (Fig. \ref{SL1_fludrop} (b)). These observations echo a similar study of the antiferromagnetic dynamics in PCMO, where the recovery time-scales span from tens of nanoseconds at low fluence to seconds at high excitation \citep{Zhou2014}. Our interpretation is that in the partially melted volume fraction the remaining order favors the relaxation to the ground state, reinforced by the still monoclinic domains. Even though the order is reduced, the symmetry is still broken for $f < f_c$, which seeds the recovery. At high excitation levels new nucleation centers have to be created before the low temperature phase can recover. Hence in analogy to supercooling, the relaxation is significantly delayed. This view is in accordance with the first-order nature of the COO transition. In equilibrium the antiferromagnetic transition is, however, second order and should thus recover more quickly. In addition the previously-observed glass-like recovery of the magnetic phase \cite{Zhou2014} is characteristic of a first order transition, contradicting again the static picture. These observations show that the antiferromagnetic state cannot grow without the pre-establishment of the charge and orbital-order. The magnetic transition is thus limited by the time the COO state needs to emerge, hence displaying the same first-order behaviour and the same delay in its recovery.

\section{Summary and conclusion}
We performed a detailed study of the structural dynamics after melting of the COO in PCMO. While the superlattice intensity related to the charge and orbital order decreases continuously with fluence, the monoclinic-to-orthorhombic transition is slower and displays a first order-like threshold behaviour. Both components of the photo-induced transition are, however, linked to the same critical excitation fluence. Comparing  these results to previous studies of the COO and magnetic dynamics, we find that all of the photo-induced dynamics are actually related to the same critical fluence $f_c$, depicting a comprehensive picture of the photo-induced transition in manganites, where the charge, structure and spin dynamics fit together.

Finally the completion of the phase transition for fluences above $f_c$ is also found to provoke a significant delay in the recovery dynamics that is well explained by the first-order nature of the thermodynamic structural transition. We also argue that the charge and orbitally-ordered phase is a necessary precursor for the establishment of the antiferromagnetic order in manganites, as the emergence of the antiferromagnetic order is limited by the recovery of the charge and orbital order.

\section{Acknowledgements}
The experiment was performed at the X05LA microXAS beamline of the Swiss Light Source at the Paul Scherrer Institut in Villigen, Switzerland. This work was supported by the Swiss National Science Foundation and its National Centers of Competence in Research MUST. We thank the microXAS beamline scientists Daniel Grolimund, Valerie Samson and Dario Ferreira for their support during the experiments and Alex Oggenfuss for his help in the experimental setup. M.N. was supported by the Japan Science and Technology Agency (JST) PRESTO (JPMJPR16R5).

\bibliography{./ref_PCMOFemto}

\end{document}